\documentclass{eas}
\usepackage{graphicx}
\begin{document}

\title{Modelling Gaia CCD pixels with Silvaco 3D engineering software} 
\runningtitle{Modelling Gaia CCD pixels}
\author{G. M. Seabroke}\address{e2v centre for electronic imaging, The Open University, Walton Hall, Milton Keynes, MK6 7AA, UK}\secondaddress{Mullard Space Science Laboratory, University College London, Holmbury St. Mary, Dorking, Surrey, RH5 6NT, UK \email{gms@mssl.ucl.ac.uk}}
\author{T. ProdÕhomme}\address{Leiden Observatory, Leiden University, The Netherlands}
\author{G. Hopkinson}\address{Surrey Satellite Technology Ltd., Sevenoaks,ÊUK}
\author{D. Burt}\address{e2v technologies plc, 106 Waterhouse Lane, Chelmsford, Essex, CM1 2QU, UK}
\author{M. Robbins}\sameaddress{5}
\author{A. Holland}\sameaddress{1}
\begin{abstract}
Gaia will only achieve its unprecedented measurement accuracy requirements with detailed calibration and correction for radiation damage.  We present our Silvaco 3D engineering software model of the Gaia CCD pixel and two of its applications for Gaia: (1) physically interpreting supplementary buried channel (SBC) capacity measurements (pocket-pumping and first pixel response) in terms of e2v manufacturing doping alignment tolerances; and (2) deriving electron densities within a charge packet as a function of the number of constituent electrons and 3D position within~the charge packet as input to microscopic models~being developed to simulate radiation damage.
\end{abstract}
\maketitle
\section{CCD pixel model}

\begin{figure}
\includegraphics[width=\textwidth]{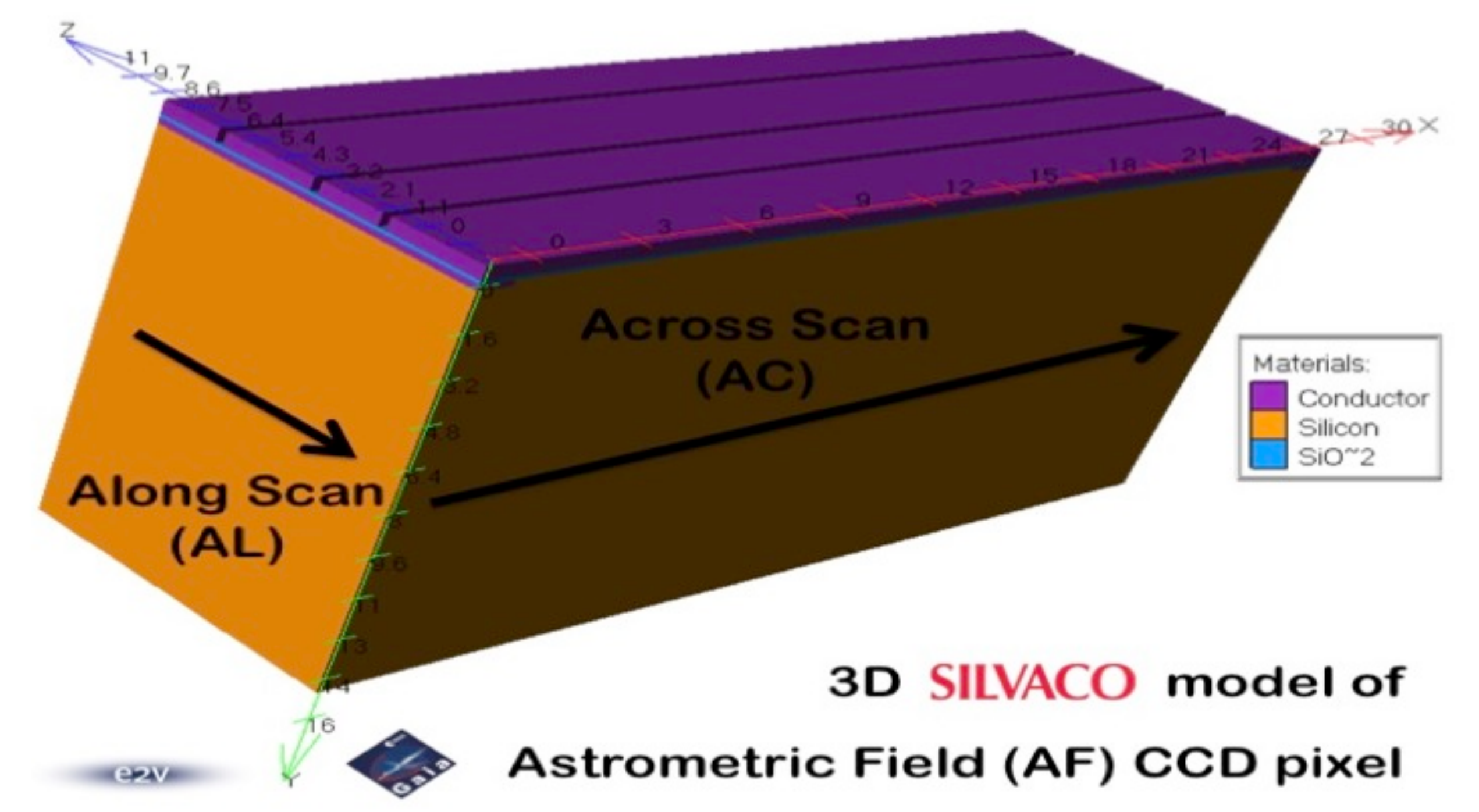}
\caption{3D Silvaco model of a single AF CCD pixel.  In the AL direction, the model has 4 electrodes that make up 1 pixel.  In the AC direction, the model runs from half way through one Anti-Blooming Drain (ABD, seen above as a conductor region beneath the oxide) to half way through the next ABD, delineating 1 CCD column.}
\end{figure}

Silvaco is a commercially available suite of engineering software.  We have been using ATLAS, Silvaco's device simulation framework: ATLAS is a physically based two and three dimensional device simulator. It predicts the electrical behaviour of specified semiconductor structures and provides insight into the internal physical mechanisms associated with device operation.  ATLAS has been successfully benchmarked against e2v measurements and other simulation software (\cite{seabroke2009}).  The 3D Silvaco model of the AF CCD pixel (see Fig. 1) consists of four different doping regions: buried channel (BC, \cite{seabroke2010a}), supplementary buried channel (SBC), ABD shielding and the ABD itself; using doping derived to agree with proprietary e2v manufacturing processes.    

\section{Interpreting measurements}

\begin{figure}
\includegraphics[width=\textwidth]{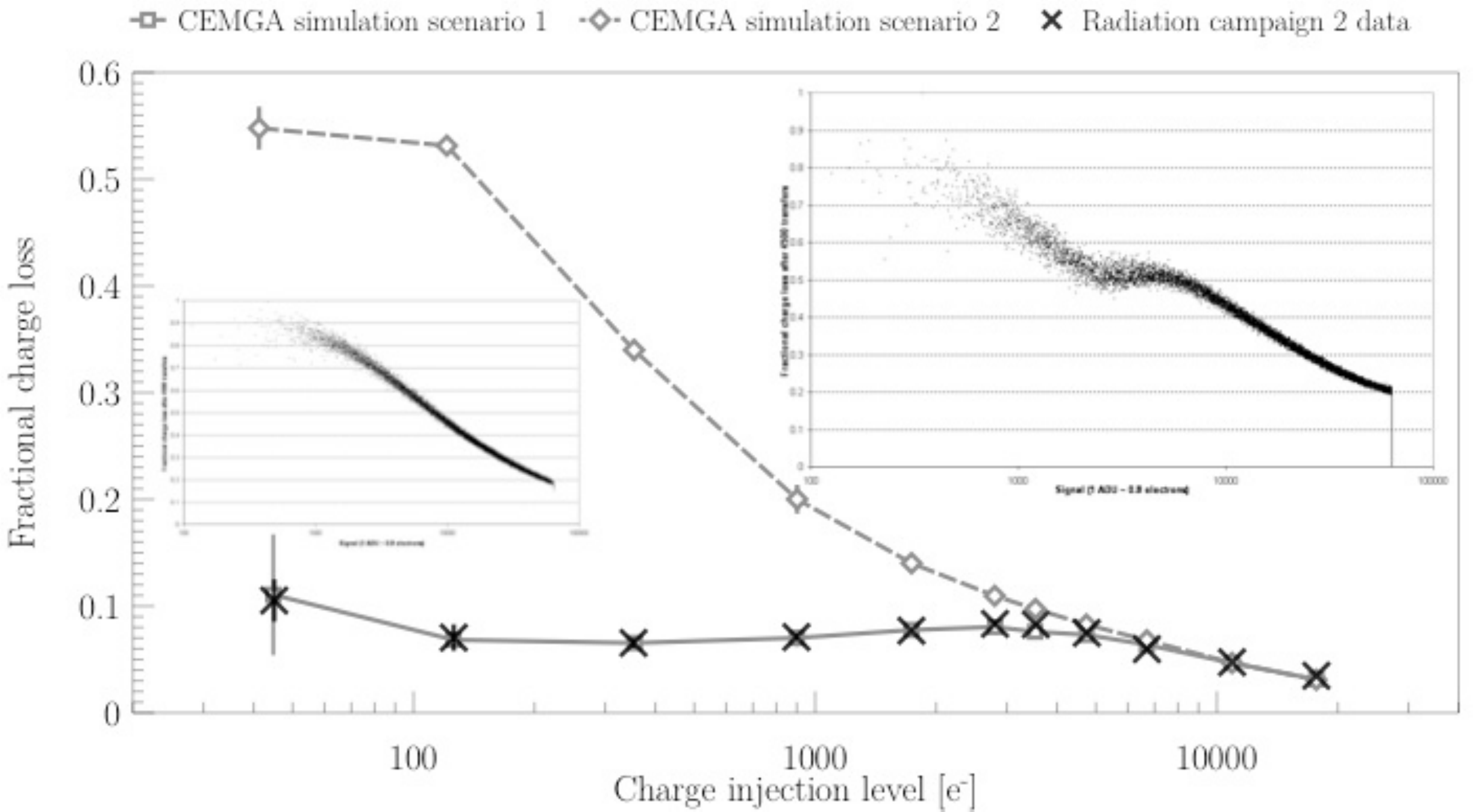}
\caption{CTI Effects Models for GAia (CEMGA) Monte Carlo simulations at the pixel level of First Pixel Response (FPR) measurements of a Gaia AF CCD: with working SBCs in both AL halves of the CCD (scenario 1) and; no SBCs in the upper AL half and working SBCs in the lower AL half (scenario 2).  In addition to the radiation campaign 2 data agreeing with the model with working SBCs in both AL halves, the insets show FPR measurements of different stitch blocks in an Engineering Model (EM) Gaia AF CCD exhibiting both scenarios.}
\end{figure}

\cite{kohley2009} first discovered the two possible SBC scenarios (see Fig. 2 for details) and consequent SBC capacities using the pocket-pumping technique on a single close-reject AF Flight Model (FM). This discovery is supported by the Fig. 2 insets. The Silvaco pixel model has explained these scenarios in terms of doping alignment on a stitch block basis (\cite{seabroke2010b}).  In the absence of pocket-pumping measurements, a comparison of FPR measurements to CEMGA simulations can distinguish between the 2 possible SBC scenarios and derive minimum SBC capacities.  In the absence of systematic SBC capacity testing of FMs prior to launch, Seabroke et al. {\it in prep.} will apply this method to 7 EM CCDs in order to increase the sample size and derive minimum SBC capacities, which can be interpreted by the Silvaco pixel model to constrain the most likely SBC scenario. 

\section{Electron densities}

\begin{figure}
\includegraphics[width=\textwidth]{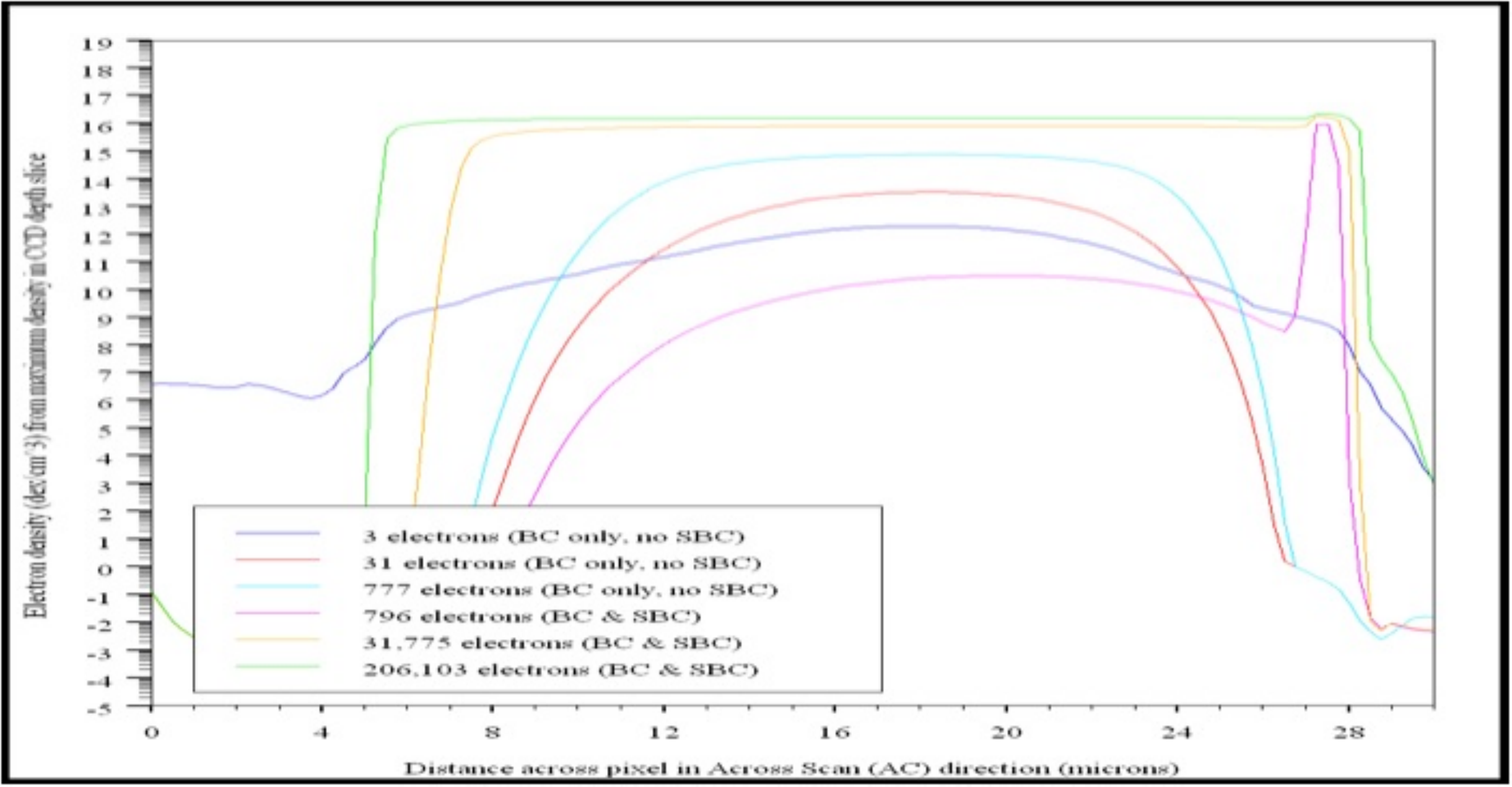}
\caption{Silvaco pixel model-simulated AC electron density profiles through different sized charge packets.  The AC cut is such that there is half a ABD at 0-1 $\mu$m and another half a ABD at 29-30 $\mu$m.  The SBC (at around 27-28 $\mu$m) causes the bimodal distribution at low signal size (when the SBC is present in the simulation) and the BC runs from 4.5 to 28 $\mu$m.}
\end{figure}

The original motivation for developing the Silvaco pixel model for Gaia was to provide electron densities of charge packets as a function of position within the pixel and as a function of the number of constituent electrons.  This will be used by CEMGA microscopic simulations to calculate how many radiation-induced traps any charge packet will meet and whether the trap will capture one of the charge packets' electrons, the probability of which depends on electron density in the vicinity of the trap (see \cite{seabroke2008} for more details).  Fig. 3 shows a representative sample of electron density profiles.  CEMGA currently uses an analytical electron density distribution (Gaussians).  Fig. 3 shows that this assumption is not physical but can be made more realistic by including a flattening factor in the analytical distribution.

\section*{Acknowledgments}
 

The Silvaco ATLAS software license fee is generously funded by the UK VEGA Gaia Data Flow System grant thanks to F. van Leeuwen and G. Gilmore.  Thanks also go to the members of the Gaia Radiation Task Force for stimulating meetings and discussions and R. Kohley for clarifying the text. 
\newpage
\bibliography{seabroke_elsa_proc_2010}   
\bibliographystyle{astron}   

\end{document}